# Dynamic disentanglement of photoflexoelectricity and flexophotovoltage

Zhiguo Wang[1,2,3], Yuanyang Guo[2], Zhenggang Rao[1,2,3], Zhibin Wen[1,2,3], Massimiliano Stengel[4,5], Longlong Shu[1,2,3]✉, Gustau Catalan[4,6]✉

[1.] *School of Physics and Materials Science, Nanchang University, Nanchang 330031, Peoples R China.*

[2.] *Jiangxi Provincial Center for Interdisciplinary Studies in Future Display, Nanchang University, Nanchang 330031, Peoples R China.*

[3.] *The Jiangxi Provincial Engineering Technology Research Center for Advanced Functional Film Materials, Nanchang University, Nanchang 330031, Peoples R China.*

[4.] *Institucio Catalana de Recerca i Estudis Avançats (ICREA), Barcelona, Catalonia.*

[5.] *Institut de Ciencia de Materials de Barcelona (ICMAB), Consejo Superior de Investigaciones Científicas (CSIC), Campus Universitat Autonoma de Barcelona, Barcelona, Catalonia.*

[6.] *Institut Catala de Nanociencia i Nanotecnologia (ICN2), CSIC and BIST, Campus UAB, Bellaterra, 08193 Barcelona, Catalonia*

✉*e-mail*: llshu@ncu.edu.cn; gustau.catalan@icn2.cat

**ABSTRACT**

The coupling between light and strain gradients shows two kinds of effects: light-enhanced flexoelectricity (photo-flexoelectricity) and gradient-enhanced photovoltage (flexophotovoltage). Although these effects originate from fundamentally different physical mechanisms (one is light-enhanced electromechanical coupling, the other is a bulk photovoltaic effect), in this article we show that dynamic flexoelectric measurements of semiconductors under illumination intrinsically contain contributions from both. To allow disentangling them, we have developed a general theoretical framework for their combined response in oscillating systems, demonstrating that the two contributions can be unambiguously separated through their distinct frequency and phase dependencies. We have validated these predictions using oscillating cantilever measurements on centrosymmetric perovskite semiconductors (SrTiO$_3$ and methyl-ammonium lead bromide, MAPbBr$_3$), obtaining self-consistent values for the coefficients both effects which are in excellent agreement with independent static measurements. Our results establish a general protocol for disentangling both light–strain-gradient couplings using only oscillatory measurements, and clarify the interpretation of flexoelectric measurements under illumination.

When a material is subjected to non-uniform stress (such as a bending deformation), the resulting strain gradient induces a deviation of the positive and negative charge centers within the crystal, resulting in a macroscopic electrical polarization. This phenomenon is known as flexoelectricity. Flexoelectricity has risen to prominence with the triple realization that (i) It can exist in materials of any and all symmetries [1,2] (ii) it can be large at the nanoscale, where large gradients are easier to achieve [3-5], and (iii) it is not exclusive to dielectrics [6-8] and can exist in semiconductors [9,10] and even metals [11-13]. The existence of flexoelectricity in semiconductors, in turn, means it can interact with the photoelectric properties of such materials, yielding flexo-photovoltaic (FPV) [14-17] and photo-flexoelectric (PFE) effects [10]. These effects are quantitatively important: the FPV and the PFE have respectively produced photovoltages bigger than the semiconductor band gap [18] and the largest flexoelectric coefficients ever recorded [10].

Although their names are similar and formally, they look like reciprocal effects, the FPV and PFE are fundamentally different. The substantive part of the FPV is a photovoltaic effect, i.e. the conversion of light into electricity. The adjective –the "flexo" prefix- indicates that this photovoltaic effect is enabled by bending: sample curvature breaks a material's centrosymmetry, thus enabling a bulk photovoltaic effect [14-17]. If the curvature is fixed during the measurement, there is no mechanical work, and thus the sole energy input is light, with the photon flow converted by the static curvature into a *direct current* [14,17]. In contrast, the PFE is an electromechanical effect: it converts a mechanical deformation (strain gradient) into electrical polarization, and it is measured as a transient displacement current generated by the *change* of polarization induced by the change in curvature. When this measurement is done under illumination, the magnitude of the current increases, and this is what we call PFE [10]. PFE measurements are therefore inherently dynamic; they derive energy from the mechanical work required to bend a material and are proportional to the rate of change of said bending [10].

While both the FPV and the PFE are sensitive to the magnitude of the strain gradient and the intensity of the illumination, only the former can generate a current under static deformation. However, under oscillatory bending, *both* the PFE and the FPV can generate an alternate current: the PFE contribution, which is proportional to the rate of change of bending, and the FPV contribution, which is proportional to the amount of bending, which oscillates. It is the purpose of this paper to evidence our prediction of coexistence of BPV and PFE in all oscillating flexoelectric measurements under illumination, and to provide a method to separate them out. We will start with a general theoretical framework, and validate the predictions of the model with oscillatory measurements under illumination for two materials SrTiO$_3$ and MAPbBr$_3$.

Let us start with the flexoelectric response of a capacitor subject to an oscillatory bending force. The change of strain gradient with time can be expressed as

$$\frac{\partial \varepsilon_{11}}{\partial x_3} = G \sin(2\pi ft) \tag{1}$$

where $G$ and $f$ are the amplitude of the curvature oscillations and the vibration frequency, respectively.

The strain gradient induces a flexoelectric polarization given by

$$P_3 = \mu_{eff} \frac{\partial \varepsilon_{11}}{\partial x_3} \tag{2}$$

where $\mu_{eff}$ is the effective flexoelectric coefficient. When the electrodes are connected in short-circuit, this polarization is compensated by a charge density at the electrode given by

$$Q = P_3 A = A\mu_{eff} \frac{\partial \varepsilon_{11}}{\partial x_3} \tag{3}$$

where $A$ is the electrode area. Since the strain gradient changes over time, so does the charge. The resulting ac current, which can be measured by an amperemeter connected in series to the capacitor, will be

$$i_{PFE} = \frac{dQ}{dt} = 2\pi f A G \mu_{eff} \cos(2\pi ft) = 2\pi f A G \mu_{eff} \sin\left(2\pi ft + \frac{\pi}{2}\right) \tag{4}$$

Flexoelectricity therefore induces an alternate current (ac) that is proportional to the frequency of the mechanical oscillation and phase-shifted 90 degrees ($\frac{\pi}{2}$ radians) with respect to the input deformation. These considerations apply equally to flexoelectric coefficients measured in the dark or with light (aka photoflexoelectricity, PFE). For the purpose of this discussion, then, $\mu_{eff}$ will be the PFE coefficient.

Let us now analyze the flexo-photovoltaic response (FPV). When a crystal is bent, a flexoelectric polarization appears which, in turn, enables a bulk photovoltaic effect. This photovoltaic effect is proportional to the polarization and therefore to the strain gradient, so it generates a photocurrent in phase with the strain gradient (or 180 degrees off, depending on the sign of the bulk photovoltaic tensor coefficients; here we will assume positive coefficients without loss of generality). The flexophotovoltaic current is thus

$$i_{FPV} = \varphi \frac{\partial \varepsilon_{11}}{\partial x_3} = \varphi A G \sin(2\pi ft) \tag{5}$$

where $\varphi$ is the flexo-photocurrent density coefficient (defined as the ratio between the amplitude of the photocurrent density, $i_{FPV}/A$, and the strain gradient). This photovoltaic current is ac and oscillates with the same frequency as the PFE. However, unlike the PFE, the FPV current is (i) independent of oscillation frequency, and (ii) in phase with the time-dependent curvature. These features can be exploited to separate the two contributions to the total current density measured by a bent capacitor under oscillation, which is given by

$$j_{AC} = \frac{i_{FPV} + i_{PFE}}{A} = \varphi G \sin(2\pi f t) + \mu_{eff} 2\pi f G \cos(2\pi f t)$$
$$= \sqrt{\varphi^2 + (2\pi f \mu_{eff})^2} G \sin(2\pi f t - \alpha)$$

(6)

where

$$\alpha = \arctan(2\pi f \mu_{eff}/\varphi) \tag{7}$$

In quasi-static mode ($f\sim 0$), $2\pi f \mu_{eff} \ll \varphi$ and $\alpha \sim 0$, so that the signal is dominated by the FPV –as expected, since in the limit f=0 the only possible source of energy is photovoltaic. With increasing oscillation frequency, however, we approach the limit $2\pi f \mu_{eff} \gg \varphi$ and $\alpha \sim \pi/2$, where the PFE dominates, reflecting the increasing contribution from the mechanical work of the cantilever oscillations. In between, the amplitude of the current as a function of frequency is a well-defined function that we can use to quantify both coefficients; specifically, the amplitude of the current density divided by the strain gradient will scale as

$$\left(\frac{j_{AC}}{G}\right)^2 = \varphi^2 + (2\pi f \mu_{eff})^2 \tag{8}$$

Therefore, measuring and plotting the amplitude ($j_{AC}/G$)$^2$ as a function of $f^{\,2}$ should provide a linear set of data where the slope is proportional to $\mu_{eff}^{\,2}$ and the intercept at the origin is $\varphi^2$. Let us test these predictions.

We have used a dynamic mechanical analysis system (DMA, DMA 850, TA instrument, USA) to drive the oscillation of a capacitor beam (the flexoelectric sample) fixed at one end, as shown in Fig. 1. In single-clamp cantilever geometry, the amplitude of the strain gradient is calculated from the vertical deflection of the free end using [19]

$$G \equiv \frac{\partial \varepsilon_{11}}{\partial x_3} = \frac{3w(L)}{L^3}\left(1 - \frac{x}{L}\right) \tag{9}$$

where *w(L)* is the vertical deflection delivered by the piezoelectric actuator at the end of the cantilever, *L* is the length of the cantilever and *x* is the horizontal position from the center of the electrode to the fixed end, respectively.

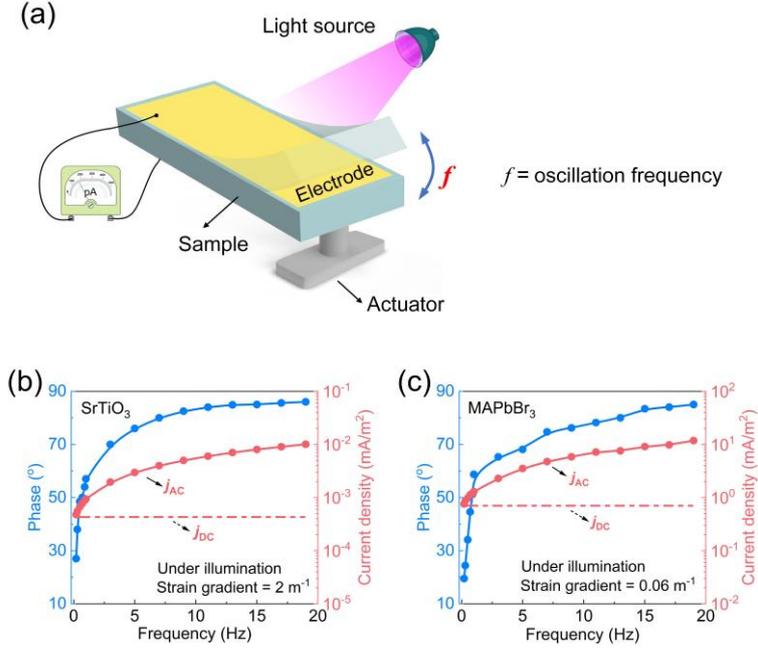

**Figure 1**. (a) Schematic diagram of the measurement setup for photoflexoelectric and flexophotovoltaic effects. An external actuator induces an oscillatory bending in a cantilever capacitor clamped at one end and illuminated from the side to induce a symmetric generation of photocarriers that then get redistributed by the oscillatory strain gradient, generating an alternate current. Measurement of the amplitude of the current density and its phase shift with respect to the input force for (b) $SrTiO_3$ and (c) $MAPbBr_3$. The frequency range is from 0.1 to 20Hz. The dotted line in the figure shows the purely photovoltaic current of the sample under static bending.

In our experimental setup, the charge output on the electrode is collected by a charge amplifier (SINOCERA, YE5852) connected to a lock-in amplifier (STANFORD RESEARCH SYSTEMS, SR830). The sample electrodes are led out by wires and connected to the charge amplifier to collect charge and convert it to voltage (1 pC/mV). The output end of the charge amplifier is connected to a lock-in amplifier to read the converted voltage value, which, multiplied by the capacitance of the beam, is the bending-induced charge, whose time derivative divided by the electrode area is the current density. The response under illumination was measured using a non-polarized LED light source (Model: HY-UV0003).

The samples examined were single crystals of strontium titanate $SrTiO_3$ (STO) and methyl-ammonium lead bromide $MAPbBr_3$ (MAPB), which are representative examples of their respective families: oxide and halide perovskites. Both materials are centrosymmetric, chosen to ensure that piezoelectric, piezo-voltaic and native bulk photovoltaic effects can be ruled out. The STO (100) crystal was commercially acquired from (TOPVENDOR, Beijing, China), and its dimensions are 5 mm × 15 mm × 0.1 mm (width × length × thickness), with electrode dimensions of 5 mm × 4 mm (width × length). The MAPB single crystals were prepared using these raw materials:

methylammonium bromide (CH3NH3Br, 99.5%, Advanced Election Technology), lead bromide (PbBr2, 99.9%, Advanced Election Technology), N,N-dimethylformamide (DMF, 99%, Energy Chemical). MAPB crystals were grown using the inverse temperature crystallization method [18]. We used 2000 mesh sandpaper to sand down the crystals to a lower thickness, followed by polishing the top and bottom surfaces with finer sandpaper (first 5000 mesh, then 10000 mesh) to achieve a mirror-smooth surface. The final dimensions of the MAPB single crystal were 5 mm × 15 mm × 1 mm and the electrodes were 5 mm × 4 mm (width × length).

In Fig. 1b-c, we show the oscillating bending current ($j_{AC}$) as a function of frequency for the two samples under illumination (red circles; the solid line is a guide to the eye). The static photocurrent density $j_{DC}$ measured under a constant mechanical load (equal to the maximum force applied in the oscillatory measurements) is represented by a red dotted line. As expected, $j_{AC}$ depends on frequency, growing with increasing frequency at one end and approaching the static limit at the other.

The phase difference between the output signal and the input vibration was also measured (blue data). We can see that at very low frequencies (0.1 Hz, close to static deformation), the phase difference between current and bending tends to 0 degrees, implying the FPV coefficient is positive; if it was negative, the phase would approach 180 degrees. In this low frequency, therefore, the FPV effect is dominant. As the oscillation frequency is increased, however, the mechanical work input grows in direct proportion to the oscillation frequency, and gradually the PFE takes over, resulting in increasing current and a gradual change of the current phase towards 90 degrees.

Eqs. 7 and 8 can be used to fit both the phase shift and the amplitude of the signals, thus providing separate measurements of the FPV coefficient ($\varphi$) and the PFE coefficient ($\mu_{eff}$). We provide these fits in Fig. 2.

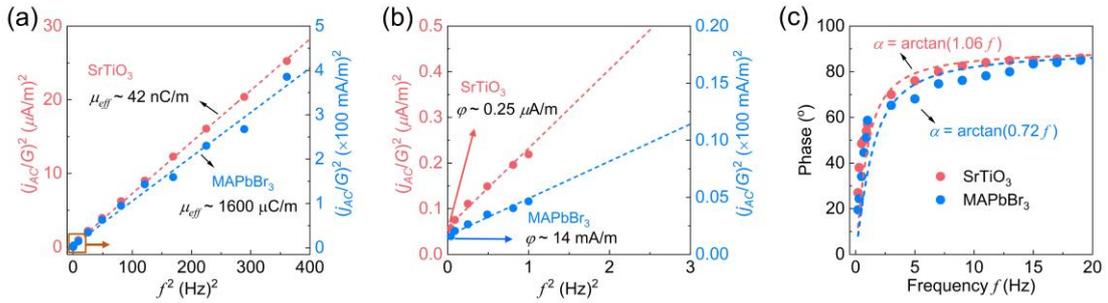

**Figure 2**. (a) Square of $j_{AC}/G$ as a function of square of the oscillation frequency for SrTiO3 and MAPbBr3 crystals, and fit of the data with a linear regression. From the slope of this fit we obtain the PFE coefficient using eq. (8). (b) Zoom in to a smaller coordinate range of panel (a), showing the intersection point of the fitted line segment and the y-axis; this intersection corresponds to the square of the FPV coefficient. (c) phase shift as a function of frequency, and fit to Equation 7.

As shown in Fig. 2a, plotting $(j_{AC}/G)^2$ as a function of the square of the frequency, $f^2$, yields a linear relationship. Fitting the data with Eq. 8, the slope is $(2\pi\mu_{eff})^2$, from which we obtain the PFE coefficient $\mu_{eff}$. For STO and MAPB, the coefficients are 42 nC/m and 1600 μC/m, respectively. If we zoom in closer to the origin (Fig. 2b), we can find the intersection with the y-axis, the value of which is the square of the FPV coefficient $(\varphi)^2$. From this value, we obtain the FPV coefficient $\varphi$; for STO and MAPB, $\varphi$ are 0.25 μA/m and 14 mA/m, respectively.

We can also access these coefficients by looking at the phase angle as a function of frequency (Fig. 2c). By fitting the data using Eq. 7, we obtain the ratio $2\pi\mu_{eff}/\varphi$ of the PFE coefficient to the FPV coefficient (Fig. 2a). For MAPB and STO, the ratios are 0.72 and 1.06, respectively. If we substitute the PFE coefficients $\mu_{eff}$ using the values obtained in Fig. 2a, from the ratio $2\pi\mu_{eff}/\varphi$ we obtain $\varphi$ of 0.24 μA/m and 13.8 mA/m, which is very close to the value obtained from the fit in Fig. 2b. The results of fitting eq. 7 and eq. 8 are therefore self-consistent. Finally, we can also compare these values of $\varphi$ with the directly measured FPV coefficients under static deformation, which are $\varphi$(STO) ~ 0.22 μA/m and $\varphi$(MAPB) ~ 12 mA/m [18], in excellent agreement with the values obtained from the dynamic measurements in the present work.

To sum up: in dynamic measurements, both the FPV and the PFE effects contribute to the total signal, with the proportion of the two depending on the frequency of the input vibration. The method presented here allows measuring accurately both the FVP and the PFE coefficients with a single experiment. The result is important to correctly describe the physics, but also to design applications, because the transducing functions of these two effects are completely different (one is photovoltaic, the other is electromechanical). We hope that the work presented here clarifies their difference and will help determining the appropriate coefficient values to best utilize bent semiconductors for energy harvesting.

**ACKNOWLEDGMENTS**

This work was supported by the National Natural Science Foundation of China (Grants No. 12404106 and 52472124) and the Natural Science Foundation of Jiangxi Province (Grants No. 20252BAC200609). ICN2 is supported by the Severo Ochoa program from Spanish MCIN / AEI (Grant No. CEX2021-001214-S. GC acknowledges support from the EU through HORIZON-MSCA-SE-3D-TOPO action (Grant agreement ID: 101236483) and from Spain's MICINN through grant PID2023-148673NB-I00 (Project GRIPHO2).

# REFERENCES

1. P. Zubko, G. Catalan, and A. K. Tagantsev, Flexoelectric effect in solids. *Annu. Rev. Mater. Res.* **43**, 387–421 (2013).
2. L. L. Shu, X. Y. Wei, T. Pang, X. Yao, and C. L. Wang, Symmetry of flexoelectric coefficients in crystalline medium. *J. Appl. Phys.* **110**, 104106 (2011).
3. M. S. Majdoub, P. Sharma, and T. Cagin, Enhanced size-dependent piezoelectricity and elasticity in nanostructures due to the flexoelectric effect. *Phys. Rev. B* **77**, 125424 (2008).
4. U. K. Bhaskar, N. Banerjee, A. Abdollahi, Z. Wang, D. G. Schlom, G. Rijnders, and G. Catalan, A flexoelectric microelectromechanical system on silicon. *Nat. Nanotechnol.* **11**, 263–266 (2016).
5. L. L. Shu, Z. G. Wang, R. H. Liang, Z. Zhang, S. W. Shu, C. X. Tang, F. Li, R.-K. Zheng, S. M. Ke, and G. Catalan, Intrinsic flexoelectricity of van der Waals epitaxial thin films. *Phys. Rev. B* **106**, 024108 (2022).
6. X. Wen, Q. Ma, A. Mannino, M. Fernandez-Serra, S. Shen, and G. Catalan, Flexoelectricity and surface ferroelectricity of water ice. *Nat. Phys.* **494**, 3935 (2025).
7. X. Wen, Q Ma, J. Liu, U. Saeed, S. Shen, and G. Catalan, Streaming flexoelectricity in saline ice. *Nat. Mater*. **137**, 1830 (2025).
8. J. W. Hong, and D. Vanderbilt, First-principles theory and calculation of flexoelectricity. *Phys. Rev. B* **88**, 174107 (2013).
9. J. Narvaez, F. Vasquez-Sancho, and G. Catalan, Enhanced flexoelectric-like response in oxide semiconductors. *Nature* **538**, 219–221 (2016).
10. L. L. Shu, S. M. Ke, L. F. Fei, W. B. Huang, Z. G. Wang, J. H. Gong, X. N. Jiang, L. Wang, F. Li, S. J. Lei, Z. G. Rao, Y. B. Zhou, R. K. Zheng, X. Yao, Y. Wang, M. Stengel, and G. Catalan, Photoflexoelectric effect in halide perovskites. *Nat. Mater.* **19**, 605-609 (2020).
11. A. Zabalo, and M. Stengel, Switching a polar metal via strain gradients. *Phys. Rev. Lett.* **126**, 127601 (2021).
12. W. Peng, S. Y. Park, C. J. Roh, J. Mun, H. Ju, J. Kim, E. K. Ko, Z. G. Liang, S. Hahn, J. F. Zhang, A. M. Sanchez, D. Walker, S. Hindmarsh, L. Si, Y. J. Jo, Y. Jo, T. H. Kim, C. Kim, L .F. Wang, M. Kim, J. S. Lee, T. W. Noh, and D. Lee, Flexoelectric polarizing and control of a ferromagnetic metal. *Nat. Phys.* **20**, 450-455 (2024).
13. G. Catalan, Metal poles around the bend. *Nat. Phys.* **20**, 358–359 (2024).
14. M.-M. Yang, D. J. Kim, and M. Alexe, Flexo-photovoltaic effect. *Science* **360**, 904–907 (2018).
15. B. W. Zhang, D. Tan, X. D. Cao, J. Y. Tian, Y. G. Wang, J. X. Zhang, Z. L. Wang, and K. L. Ren, Flexoelectricity-enhanced photovoltaic effect in self-polarized flexible PZT nanowire array devices. *ACS Nano* **16**, 7834–7847 (2022).

16. Z. G. Wang, H. Q. Zhong, Z. Y. Liu, X. T. Hu, L. L. Shu, and G. Catalan, Strain-gradient-induced modulation of photovoltaic efficiency. *Matter* **8**, 101930 (2025).

17. Z. Z. Jiang, Z. Y. Xu, Z. N. Xi, Y. H. Yang, M. Wu, Y. K. Li, X. Li, Q. Y. Wang, C. Li, D. Wu, and Z. Wen, Flexoelectric-induced photovoltaic effects and tunable photocurrents in flexible LaFeO$_3$ epitaxial heterostructures. *J. Materiomics* **8**, 281-287 (2022).

18. Z. G. Wang, S. W. Shu, X. Y. Wei, R. H. Liang, S. M. Ke, L. L. Shu, and G. Catalan, Flexophotovoltaic effect and above-bandgap photovoltage induced by strain gradients in halide perovskites. *Phys. Rev. Lett.* **132**, 086902 (2024).

19. Z. G. Wang, C. C. Li, H. Y. Xie, Z. Zhang, W. B. Huang, S. M. Ke, and L. L. Shu, Effect of grain size on flexoelectricity. *Phys. Rev. Appl.* **18,** 064017 (2022).